\documentclass{article}
\usepackage{spconf,amsmath,graphicx}

\usepackage{subfigure} 
\graphicspath{{./image_new/}}

\usepackage{amssymb}

\usepackage{epstopdf}
\usepackage{tabularx}
\usepackage{multirow}
\usepackage{textcomp}
\usepackage{url}
\usepackage{lipsum}
\usepackage{makecell}

\usepackage{bm}
\usepackage{calc}
\usepackage{cite}
\usepackage[table]{xcolor}

\definecolor{color2}{HTML}{eff3ff}
\definecolor{color4}{HTML}{c6dbef}
\definecolor{color8}{HTML}{9ecae1}
\definecolor{color16}{HTML}{6baed6}

\usepackage{hhline}

\usepackage{color}
\usepackage{dsfont}

\usepackage{algorithm}
\usepackage[noend]{algpseudocode}

\title{Polyphonic audio event detection: \\ multi-label or multi-class multi-task classification problem?}
\name{\begin{tabular}{c}Huy Phan$^{*1,2}$, Thi Ngoc Tho Nguyen$^{3}$, Philipp Koch$^{4}$, Alfred Mertins$^{4}$
\end{tabular}}

\address{$^1$Centre for Digital Music, Queen Mary University of London, UK \\ $^2$The Alan Turing Institute, UK \\ $^3$School of Electrical and Electronic Engineering, Nanyang Technological University, Singapore \\
	$^4$Institute for Signal Processing, University of L\"ubeck, Germany \\
	{$^\ast$Correspondence email: \tt h.phan@qmul.ac.uk} 
}

\begin{document}
	\ninept
	\maketitle
	\begin{abstract}
		Polyphonic events are the main error source of audio event detection (AED) systems. In deep-learning context, the most common approach to deal with event overlaps is to treat the AED task as a multi-label classification problem. By doing this, we inherently consider multiple one-vs.-rest classification problems, which are jointly solved by a single (i.e. shared) network. In this work, to better handle polyphonic mixtures, we propose to frame the task as a multi-class classification problem by considering each possible label combination as one class. To circumvent the large number of arising classes due to combinatorial explosion, we divide the event categories into multiple groups and construct a multi-task problem in a divide-and-conquer fashion, where each of the tasks is a multi-class classification problem. A network architecture is then devised for multi-class multi-task modelling. The network is composed of a backbone subnet and multiple task-specific subnets. The task-specific subnets are designed to learn time-frequency and channel attention masks to extract features for the task at hand from the common feature maps learned by the backbone. Experiments on the TUT-SED-Synthetic-2016 with high degree of event overlap show that the proposed approach results in more favorable performance than the common multi-label approach.
	\end{abstract}
	\vspace{-0.15cm}
	\section{Introduction}
	\vspace{-0.15cm}
	
	During the last five years, the research community has witnessed significant progress in audio event detection (AED) \cite{Stowell2015, Mesaros2017, Phan2015, McLoughlin2017} (and localization \cite{He2021,Adavanne2019,Cao2021,Nguyen2021b,Shimada2021}), which has benefited from the evolution of deep-learning field and the initiative of the annual detection and classification of acoustic scenes and events (DCASE) challenge
	\cite{Mesaros2017,Mesaros2017b,Politis2020}. However, polyphonic events, i.e. when more than one event instances happen at the same time, still remain as a major challenge for AED. Majority of existing works technically aimed to improve overall detection performance regardless of monophony or polyphony.
	Such as system often biases toward dominant monophonic audio events in the training data and performs less well for under-present polyphonic audio events.
	In practice, these systems have been shown to work well on monophonic events while polyphonic events remain a prominent source of detection errors \cite{Nguyen2021, Cakir2017, Phan2019b}, especially when the number of events in a mixture is large.
	
	\begin{figure*} [!t]
		\centering
		\includegraphics[width=0.715\linewidth]{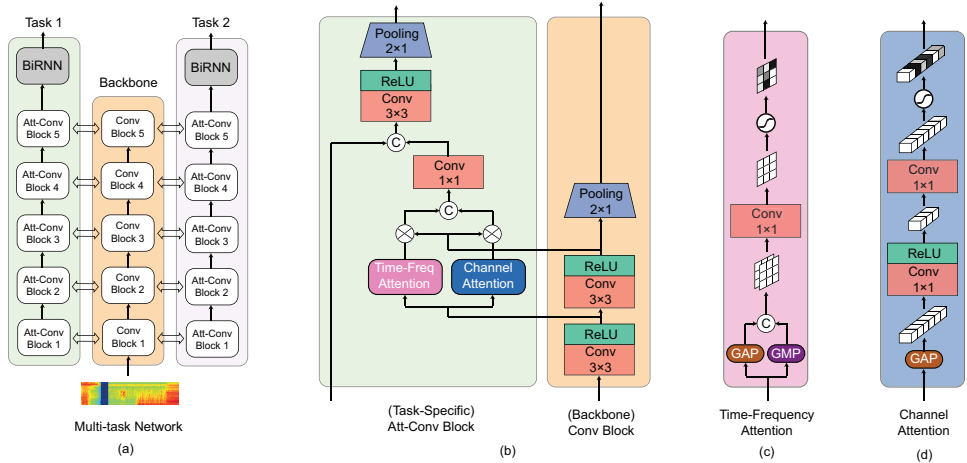}
		\vspace{-0.4cm}
		\caption{(a) Illustration for overall structure of the proposed multi-task network architecture with a backbone subnet (middle) and two task-specific subnets (left and right). (b) Architecture of the backbone's convolutional block (right) and a task-specific subnet's attention-convolution block (left). (c) Architecture of the TF attention component. (d) Architecture of the channel attention component.}
		\label{fig:architecture}
		\vspace{-0.2cm}
	\end{figure*}
	
	In existing works, in order to handle event overlaps, the AED task is often framed as a multi-label classification problem, which is then solved by a deep neural network 
	with multi-label output format
	\cite{Cakir2017, jung2019sed, kong2020weaklysed, Mesaros2021SoundTutorial}. In essence, the single network is tasked to jointly deal with multiple one-vs.-rest binary classification problems at once. Doing so is arguably suboptimal. First, the network cannot capture the interaction between different target event categories. Second, an event mixture is processed in the same way as a standalone event instance in the processing pipeline. As a result, the hard parameter sharing network is forced to learn task-shared generalizable representation, but lacks of ability to learn desired task-specific features tailored to each task. In \cite{Cao2021}, Cao \emph{et al.} showed that a multi-task network with event-independent prediction tracks led to improved performance over the common multi-label treatment.
	
	In this work, we propose a new approach to deal with polyphonic events specifically. Instead of the common multi-label setting, we consider all possible event mixtures (e.g. combinations of different event categories), where one of such combinations is treated as a target class to be determined by a deep neural network. To circumvent the combinatorial explosion problem, when the number of event categories is large (e.g. 16 target event categories will result in $2^{16}\!=\!65,536$ classes), we divide the event categories into multiple subsets, and thus, form a multi-task problem. Each subset is considered as a task that is a multi-class classification problem on its own. A network architecture is then devised for multi-class multi-task modelling ~\cite{vandenhende2020mtlsurvey, crawshaw2020mtlsurvey}. The network is composed of a backbone subnet and multiple task-specific subnets. On the one hand, the backbone subnet is expected to learn task-shared generalizable features. On the other hand, the task-specific subnets, one for each designated modelling task, make use of time-frequency (TF) attention layers, channel attention layers, and task-specific convolutional layers. The attention layers are to extract task-specific features from the shared feature maps learned by the backbone subnet while the  task-specific convolutional layers are supposed to refine these features for the task at hand. Experiments on the TUT-SED-Synthetic-2016 database \cite{Cakir2017} with heavy event overlaps show that the proposed approach results in better performance than the common multi-label approach, leading to improvement of up to $3.7\%$ absolute on F1-score. The improvement is seen across different overlapping degrees, monophonic events included, and over most of the event categories.

	\vspace{-0.2cm}
	\section{Multi-task decomposition}
	\label{sec:multitaskdecomposition}
	\vspace{-0.15cm}
	
	Let us assume an AED task with $Y$ target event categories. Consider all possible event mixtures, each of them becomes a class in our multi-class classification formulation. Thus, there are $Y'\!=\!2^Y$ classes in total. In other words, $Y'$ increases exponentially with respect to $Y$ (i.e. combinatorial explosion), resulting in a challenging modelling problem, when $Y$ is large if considered as is. Apart from computational overhead due to the large number of classes, we would need a lot of data with the presence of all possible mixtures in order to train a deep neural network. However, such a network would be unlikely to perform well. The rationale is that there are unlimited ways the event instances can overlap together, making the features learned from the event mixtures in the training data not generalizing well to unseen mixtures in the test data \cite{Mesaros2021SoundTutorial}.
	
	To remedy the combinatorial explosion problem, we decompose the original event categories into $N$ non-overlapping groups, $2\!\le\!N\!\le\!Y$, with $\{Y_1, Y_2, \ldots, Y_N\}$ categories in the groups, where $Y_1+Y_2+\ldots+Y_N\!=\!Y$. We then treat the detection of the event categories in one group as a task, and thus decompose the original AED task into $N$ subtasks. In addition, considering all possible event mixtures in principle of the multi-class classification formulation, the number of classes of the subtasks is  $\{2^{Y_1}, 2^{Y_2}, \ldots, 2^{Y_N}\}$. It should be noted that there is a huge number of possible ways for multi-task decomposition depending on the number of tasks $N$ and which event categories are assigned to a task. However, as long as a decomposition is given, the original problem can be addressed via the proxy of the resulting multi-task problem, which can be accomplished by, for example, a multi-task model. 
	
	\vspace{-0.2cm}
	\section{Multi-task network architecture}
	\label{sec:network}
	\vspace{-0.2cm}
	Given a multi-task decomposition with $N$  tasks as described in Section \ref{sec:multitaskdecomposition}, inspired by \cite{Liu2019b}, we propose a network architecture to deal with multi-task modelling here. As illustrated in Fig.~\ref{fig:architecture} (a), at the high level, the network is composed of one backbone subnet and $N$ subnets, one for each task. The backbone subnet is the only component directly ingesting the spectrogram input. It is supposed to learn task-shared hierarchical features along the depth of its architecture. At each level of the network hierarchy, a task-specific subnet is supposed to selectively extract features from the joint feature maps learned by the backbone via TF and channel attention masks. The extracted task-specific features are then further refined by the task-specific subnet's convolutional layers. Of note, the backbone subnet is not directly associated with any modelling task.
	
	\vspace{-0.15cm}
	\subsection{The backbone subnet}
	\label{ssec:backborne}
	\vspace{-0.15cm}
	The backbone subnet consists of five convolutional blocks (Conv-Block), as illustrated in Fig.~\ref{fig:architecture}(a). Each block is composed of two convolutional layers 
	followed by a max pooling layer. In addition, batch normalization, ReLU activation, and dropout are applied after each convolutional layer. A common filter size $3\!\times\!3$ and stride $1\!\times\!1$  are used for both convolutional layers. The max-pooling layer has its kernel size set to $1\!\times\!2$ to reduce the frequency resolution by two while keeping the time resolution unchanged. The numbers of convolutional filters in the five Conv-Block blocks are configured as $\{64, 64, 128, 128, 256\}$, respectively. Given the spectrogram input $\mathbf{S} \in \mathbb{R}^{T\times F}$ of $T$ time frames and $F$ frequency bins, the resulting feature maps of the five Conv-Block blocks are of size $\{T\!\times\!\frac{F}{2}\!\times\!64, T\!\times\!\frac{F}{4}\!\times\!64, T\!\times\!\frac{F}{8}\!\times\!128, T\!\times\!\frac{F}{16}\!\times\!128, T\!\times\!\frac{F}{32}\!\times\!256\}$, respectively.
	
	\vspace{-0.15cm}
	\subsection{The task-specific subnets}
	\label{ssec:taskspecific}
	\vspace{-0.15cm}
	Each task-specific subnet has five attention-convolution blocks (Att-Conv-Block) corresponding to the backbone's five Conv-Blocks in the network hierarchy. As illustrated in Fig. \ref{fig:architecture} (b), at a certain level of the network hierarchy, an Att-Conv-Block (in the left panel) receives its predecessor's output as input (except for the first Att-Conv-Block) and at the same time interacts with the backbone's Conv-Block (in the right panel) via TF and channel attention layers. Let $\mathbf{M}^{\text{bb}}_1$ and  $\mathbf{M}^{\text{bb}}_2$ denote the feature maps outputted by the two convolutional layers of the backbone's Conv-Block. $\mathbf{M}^{\text{bb}}_1$ and  $\mathbf{M}^{\text{bb}}_2$ have the same size owing to zero padding (also known as ``SAME'' padding) and their common configuration (cf. Section \ref{ssec:backborne}). Without lack of generalization, assume their size to be $T'\!\times\!F'\!\times\!C'$, where $T'$, $F'$, and $C'$ denote the size of the time, frequency, and channel dimension, respectively. The TF and channel attention components of the Att-Conv-Block are devised to produce a TF attention mask $\mathbf{m}_{\text{tf}}$ and a channel attention mask $\mathbf{m}_{\text{c}}$ from $\mathbf{M}^{\text{bb}}_1$. The attention masks are then applied to $\mathbf{M}^{\text{bb}}_2$ to recalibrate and separate the task-specific features out of $\mathbf{M}^{\text{bb}}_2$.
	
	{\bf Time-frequency attention.} The TF attention component is illustrated in Fig. \ref{fig:architecture} (c). It is tasked to capture the time and frequency dependencies of the convolutional feature map $\mathbf{M}^{\text{bb}}_1$. 
	To this end, global average pooling (GAP) and global max pooling (GMP) are independently applied along the channel dimension. The outputs of the two pooling operations are then concatenated along the channel dimension, yielding a tensor $\mathbf{p} \in \mathbb{R}^{T' \times F' \times 2}$. Afterwards, a $1\!\times\!1$ convolution is applied to $\mathbf{p}$, followed by sigmoid activation to produce the TF attention mask $\mathbf{m}_{\text{tf}} \in \mathbb{R}^{T' \times F' \times 1}$. This is similar to the spatial attention presented in \cite{woo2018cbam, Zamir2020}.
	
	{\bf Channel attention.}  As illustrated in Fig. \ref{fig:architecture} (d), the channel attention component is to exploit the relationship across the channels of the feature map $\mathbf{M}^{\text{bb}}_1$. GAP is first applied across the time and frequency dimensions to obtain a tensor $\mathbf{q} \in \mathbb{R}^{1 \times 1 \times C'}$. The channel dimension of $\mathbf{q}$ is then squeezed from $C'$ to $\frac{C'}{2}$ via a $1\!\times\!1$ convolution layer with $\frac{C'}{2}$ kernels and ReLU activation, followed by excitation via another $1\!\times\!1$ convolution layer with $C'$ kernels to restore the initial shape. This is similar to the idea of the squeeze-and-excitation networks proposed in \cite{Hu2018} and the channel attention module in \cite{woo2018cbam, Zamir2020}. Finally, sigmoid activation is applied to produce the channel attention mask $\mathbf{m}_{\text{c}} \in \mathbb{R}^{1 \times 1 \times C'}$. 
	
	Using the attention masks $\mathbf{m}_{\text{tf}}$ and $\mathbf{m}_{\text{c}}$, the task-specific feature maps are derived from $\mathbf{M}^{\text{bb}}_{2}$ as 
	\begin{align}
	\mathbf{M}^{*} = (\mathbf{m}_{\text{tf}}\otimes\mathbf{M}^{\text{bb}}_{2}) \oplus (\mathbf{m}_{\text{c}}\otimes\mathbf{M}^{\text{bb}}_{2}),
	\label{eq:taskspecific}
	\end{align}
	where $\otimes$ denotes an element-wise multiplication and broadcasting operator while $\oplus$ denotes concatenation along the channel dimension. Thus, $\mathbf{M}^{*}$ is of size $T'\!\!\times\!\!F'\!\!\times\!2C'$, which is then reduced to $T'\!\!\times\!\!F'\!\!\times\!\!C'$ via a $1\!\times\!1$ convolutional layer with $C'$ kernels. $\mathbf{M}^{*}$ is then concatenated to the preceding Att-Conv-Block's output along the channel dimention. The resulting feature map is further processed by a convolutional layer. This convolutional layer is expected to learn features specific to the task at hand from $\mathbf{M}^{*}$. It is configured similar to the two convolutional layers in the backborn Conv-Block (i.e., $3\!\times\!3$ filter size, $1\!\times\!1$ stride, and the same number of filters) and followed by batch normalization, ReLU activation, and dropout. Finally, a pooling layer with $1\!\times\!2$  kernel is employed to halve the frequency dimension of the feature map and make its size compatible to the computation in the subsequent Att-Conv-Block.
	
	Different from the backbone, which is not associated with any modelling task, the task-specific subnet makes use of a bidirectional recurrent neural network (BiRNN) after the last Att-Conv-Block. It is used to exploit the context information of the entire sequence to enrich the features at each time index before the classification takes place. Here, the BiRNN is realized by Gated Recurrent Units (GRU) \cite{Cho2014}, whose hidden state vectors have $H=256$ units. For classification purpose, two time-distributed fully-connected (FC) layers with 512 units each and ReLU activation are used, followed by an output layer with softmax activation. Due to the multi-class setting, categorical cross-entropy loss is used in each task-specific subnet during training. The entire multi-task network is trained to minimize the average cross-entropy loss over all the task-specific subnets.
	\vspace{-0.15cm}
	
	\section{Experiments}
	\label{sec:experiment}
	\vspace{-0.15cm}
	
	\subsection{Dataset}
	\vspace{-0.15cm}
	We employed the TUT-SED-Synthetic-2016 \cite{Cakir2017} in this study. The database was specifically designed for studying polyphonic audio event detection. It consists of 100 polyphonic mixtures created by mixing 994 monophonic event instances of 16 event categories listed on the left most column of Table \ref{tab:multitaskdecomposition}. The mixing procedure was conducted in such a way that event overlaps happen very often and the maximum degree of overlap was 6, i.e. 6 different event instances occurred at the same time. The total length of the data is 566 minutes. Further details on the dataset creation procedure can be found in \cite{Cakir2017}. Out of 100 created mixtures, 60 were used for training, 20 for evaluation, and 20 for validation. 
	\vspace{-0.15cm}
	
	\subsection{Parameters}
	\vspace{-0.15cm}
	The audio recordings were sampled at 44100 Hz. Each of them was transformed into a log Melscale spectrogram using $F\!=\!64$ Mel-scale filters in the frequency range of [50, 22050] Hz. A frame size of 40 ms with 50\% overlap was used for this transformation. We used audio segments of length $T=128$ frames as network inputs. During training, the audio segments were densely sampled (i.e., $T-1$ frame overlap) from the training recordings. However, during testing, they were sampled from the test recording without overlap. A network was trained with minibatch size of 32 for 10 epochs. Adam optimizer \cite{Kingma2015} with a learning rate of $10^{-4}$ was used for training. For regularization, a dropout rate of 0.25 was applied to the convolutional, recurrent, and FC layers. 
	
	\setlength\tabcolsep{2.5pt} 
	\begin{table}[!t]
		\caption{Multi-task decompositions given the set of event categories in the TUT-SED-Synthetic-2016 database \cite{Cakir2017}. In each column, the tasks are separated by the horizontal lines.}
		\vspace{-0.15cm}
		\footnotesize
		\begin{center}
			\begin{tabular}{|>{\arraybackslash}m{1.1in}|>{\centering\arraybackslash}m{0.4in}|>{\centering\arraybackslash}m{0.4in}|>{\centering\arraybackslash}m{0.4in}|>{\centering\arraybackslash}m{0.485in}|>{\centering\arraybackslash}m{0in} @{}m{0pt}@{}}
				\cline{1-5}
				\multirow{2}{*}{Event categories} &  \multicolumn{4}{c|}{Multi-task} & \parbox{0pt}{\rule{0pt}{0ex+\baselineskip}} \\ [-0.2ex]  	
				\cline{2-5}
				
				& 2 tasks & 4 tasks & 8 tasks & 16 tasks \parbox{0pt}{\rule{0pt}{0ex+\baselineskip}} \\ [-0.2ex] 
				\hhline{|-|----|}
				alarms \& sirens (as) & \cellcolor{color2} as & \cellcolor{color4} as & \cellcolor{color8}  as & \cellcolor{color16} as & \parbox{0pt}{\rule{0pt}{\baselineskip}} \\ [-0.2ex]  	
				\hhline{|~|>{\arrayrulecolor{color2}}->{\arrayrulecolor{black}}|>{\arrayrulecolor{color4}}->{\arrayrulecolor{black}}|>{\arrayrulecolor{color8}}->{\arrayrulecolor{black}}|-|}
				baby crying (bc)& \cellcolor{color2}bc & \cellcolor{color4} bc & \cellcolor{color8}bc & \cellcolor{color16}bc & \parbox{0pt}{\rule{0pt}{\baselineskip}} \\ [-0.2ex]  	
				\hhline{|~|>{\arrayrulecolor{color2}}->{\arrayrulecolor{black}}|>{\arrayrulecolor{color4}}->{\arrayrulecolor{black}}|-|-|}
				bird singing (bs)& \cellcolor{color2}bs & \cellcolor{color4} bs & \cellcolor{color8}bs & \cellcolor{color16}bs & \parbox{0pt}{\rule{0pt}{\baselineskip}} \\ [-0.2ex]  
				\hhline{|~|>{\arrayrulecolor{color2}}->{\arrayrulecolor{black}}|>{\arrayrulecolor{color4}}->{\arrayrulecolor{black}}|>{\arrayrulecolor{color8}}->{\arrayrulecolor{black}}|-|}
				bus~~~~~~~~~&  \cellcolor{color2}bus & \cellcolor{color4} bus & \cellcolor{color8}bus & \cellcolor{color16}bus &  \parbox{0pt}{\rule{0pt}{\baselineskip}} \\ [-0.2ex]  		
				\hhline{|~|>{\arrayrulecolor{color2}}->{\arrayrulecolor{black}}|-|-|-|}
				cat meowing (cm)&  \cellcolor{color2}cm & \cellcolor{color4} cm & \cellcolor{color8}cm & \cellcolor{color16}cm &  \parbox{0pt}{\rule{0pt}{\baselineskip}} \\ [-0.2ex]  	
				\hhline{|~|>{\arrayrulecolor{color2}}->{\arrayrulecolor{black}}|>{\arrayrulecolor{color4}}->{\arrayrulecolor{black}}|>{\arrayrulecolor{color8}}->{\arrayrulecolor{black}}|-|}
				crowd applause (ca)&  \cellcolor{color2}ca & \cellcolor{color4} ca & \cellcolor{color8}ca & \cellcolor{color16}ca & \parbox{0pt}{\rule{0pt}{\baselineskip}} \\ [-0.2ex]  	
				\hhline{|~|>{\arrayrulecolor{color2}}->{\arrayrulecolor{black}}|>{\arrayrulecolor{color4}}->{\arrayrulecolor{black}}|-|-|}
				crowd cheering (cc) & \cellcolor{color2} cc & \cellcolor{color4} cc & \cellcolor{color8}cc & \cellcolor{color16}cc & \parbox{0pt}{\rule{0pt}{\baselineskip}} \\ [-0.2ex]  		
				\hhline{|~|>{\arrayrulecolor{color2}}->{\arrayrulecolor{black}}|>{\arrayrulecolor{color4}}->{\arrayrulecolor{black}}|>{\arrayrulecolor{color8}}->{\arrayrulecolor{black}}|-|}
				dog barking (db) & \cellcolor{color2} db & \cellcolor{color4} db & \cellcolor{color8}db & \cellcolor{color16}db & \parbox{0pt}{\rule{0pt}{\baselineskip}} \\ [-0.2ex]  		
				\hhline{|~|----|}
				footsteps (fs) & \cellcolor{color2} fs & \cellcolor{color4} fs & \cellcolor{color8}fs & \cellcolor{color16}fs & \parbox{0pt}{\rule{0pt}{\baselineskip}} \\ [-0.2ex]  	
				\hhline{|~|>{\arrayrulecolor{color2}}->{\arrayrulecolor{black}}|>{\arrayrulecolor{color4}}->{\arrayrulecolor{black}}|>{\arrayrulecolor{color8}}->{\arrayrulecolor{black}}|-|}
				glass smash (gs) & \cellcolor{color2} gs & \cellcolor{color4} gs & \cellcolor{color8}gs & \cellcolor{color16}gs & \parbox{0pt}{\rule{0pt}{\baselineskip}} \\ [-0.2ex]  	
				\hhline{|~|>{\arrayrulecolor{color2}}->{\arrayrulecolor{black}}|>{\arrayrulecolor{color4}}->{\arrayrulecolor{black}}|-|-|}
				gun shot (gsh) & \cellcolor{color2} gsh &\cellcolor{color4}  gsh & \cellcolor{color8}gsh & \cellcolor{color16}gsh & \parbox{0pt}{\rule{0pt}{\baselineskip}} \\ [-0.2ex]  		
				\hhline{|~|>{\arrayrulecolor{color2}}->{\arrayrulecolor{black}}|>{\arrayrulecolor{color4}}->{\arrayrulecolor{black}}|>{\arrayrulecolor{color8}}->{\arrayrulecolor{black}}|-|}
				horsewalk (hw) & \cellcolor{color2} hw & \cellcolor{color4} hw & \cellcolor{color8}hw & \cellcolor{color16}hw & \parbox{0pt}{\rule{0pt}{\baselineskip}} \\ [-0.2ex]  	
				\hhline{|~|>{\arrayrulecolor{color2}}->{\arrayrulecolor{black}}|-|-|-|}
				mixer (mx) & \cellcolor{color2} mx & \cellcolor{color4} mx & \cellcolor{color8}mx & \cellcolor{color16}mx & \parbox{0pt}{\rule{0pt}{\baselineskip}} \\ [-0.2ex]  	
				\hhline{|~|>{\arrayrulecolor{color2}}->{\arrayrulecolor{black}}|>{\arrayrulecolor{color4}}->{\arrayrulecolor{black}}|>{\arrayrulecolor{color8}}->{\arrayrulecolor{black}}|-|}
				motorcycle (mc) &\cellcolor{color2} mc &\cellcolor{color4}  mc & \cellcolor{color8}mc & \cellcolor{color16}mc & \parbox{0pt}{\rule{0pt}{\baselineskip}} \\ [-0.2ex]   		
				\hhline{|~|>{\arrayrulecolor{color2}}->{\arrayrulecolor{black}}|>{\arrayrulecolor{color4}}->{\arrayrulecolor{black}}|-|-|}
				rain & \cellcolor{color2}rain &\cellcolor{color4}  rain & \cellcolor{color8}rain & \cellcolor{color16}rain & \parbox{0pt}{\rule{0pt}{\baselineskip}} \\ [-0.2ex]  
				\hhline{|~|>{\arrayrulecolor{color2}}->{\arrayrulecolor{black}}|>{\arrayrulecolor{color4}}->{\arrayrulecolor{black}}|>{\arrayrulecolor{color8}}->{\arrayrulecolor{black}}|-|}
				thunder (td) & \cellcolor{color2} td & \cellcolor{color4} td &\cellcolor{color8} td & \cellcolor{color16}td & \parbox{0pt}{\rule{0pt}{\baselineskip}} \\ [-0.2ex]  	
				\cline{1-5}
			\end{tabular}
		\end{center}
		\label{tab:multitaskdecomposition}
		\vspace{-0.5cm} 
	\end{table}

	\vspace{-0.15cm}
	\subsection{Baseline}
	\vspace{-0.15cm}
	For comparison, we implemented a multi-label  baseline based on convolutional recurrent neural network (CRNN), which is the most commonly used architecture for AED. The baseline's CNN part utilized the backbone subnet (cf. Section \ref{ssec:backborne}) of the proposed multi-task network. The RNN part was a GRU-based BiRNN with hidden-state size of 256. For classification purpose, two time-distributed FC layers with 512 units each and ReLU activation were used, followed by an output layer with sigmoid activation. The baseline was trained with sigmoid cross-entropy loss for multi-label classification. 
	
	\vspace{-0.15cm}
	\subsection{Evaluation metrics}
	\vspace{-0.15cm}
	We evaluated the AED performance using the frame-based F1 score as in \cite{Cakir2017}. The F1 score is the harmonic mean of precision (P) and recall (R), which are calculated as follows
	\begin{equation}
	P = \frac{TP}{TP + FP}  \quad  R = \frac{TP}{TP+FN},
	\end{equation}
	where TP, FP, and FN are the numbers of true positive, false positive, and false negative, respectively. 
	
	\setlength\tabcolsep{2.5pt} 
	\begin{table}[!t]
		\caption{The F1-scores obtained by the proposed multi-task approach and the multi-label baseline. Bold-face indicates the multi-task networks outperform the baseline.}
		\footnotesize
		\vspace{-0.2cm}
		\begin{center}
			\begin{tabular}{|>{\arraybackslash}m{0.8in}|>{\centering\arraybackslash}m{0.4in}|>{\centering\arraybackslash}m{0.385in}|>{\centering\arraybackslash}m{0.385in}|>{\centering\arraybackslash}m{0.385in}|>{\centering\arraybackslash}m{0.45in}|>{\centering\arraybackslash}m{0in} @{}m{0pt}@{}}
				\cline{1-6}
				\multirow{2}{*}{Event type} & \multirow{2}{*}{\makecell{Multi\\-label}}  &  \multicolumn{4}{c|}{Multi-task} & \parbox{0pt}{\rule{0pt}{0ex+\baselineskip}} \\ [-0.2ex]  	
				\cline{3-6}
				& & 2 tasks & 4 tasks & 8 tasks & 16 tasks & \parbox{0pt}{\rule{0pt}{0ex+\baselineskip}} \\ [-0.2ex] 
				\cline{1-6}
				alarms \& sirens & $65.1$ & $\bm{68.4}$ & $\bm{70.8}$ & $\bm{73.2}$ & $\bm{67.0}$ & \parbox{0pt}{\rule{0pt}{\baselineskip}} \\ [-0.2ex]  	
				baby crying~~~~& $52.3$ & $\bm{56.5}$ & $49.1$ & $\bm{54.8}$ & $\bm{55.3}$ & \parbox{0pt}{\rule{0pt}{\baselineskip}} \\ [-0.2ex]  	
				bird singing~~~~& $49.4$ & $\bm{50.5}$ & $\bm{52.0}$ & $48.3$ & $45.8$ & \parbox{0pt}{\rule{0pt}{\baselineskip}} \\ [-0.2ex]  
				bus~~~~~~~~~& $53.7$ & $\bm{53.9}$ & $\bm{58.4}$ & $\bm{65.0}$ & $\bm{58.8}$ &  \parbox{0pt}{\rule{0pt}{\baselineskip}} \\ [-0.2ex]  		
				cat meowing~~~& $28.4$ & $\bm{46.8}$ & $\bm{49.0}$ & $\bm{44.1}$ & $\bm{48.0}$ &  \parbox{0pt}{\rule{0pt}{\baselineskip}} \\ [-0.2ex]  	
				crowd applause~& $70.9$ & $70.2$ & $\bm{73.5}$ & $\bm{73.1}$ & $\bm{73.3}$ & \parbox{0pt}{\rule{0pt}{\baselineskip}} \\ [-0.2ex]  	
				crowd cheering~ & $69.6$ & $\bm{73.9}$ & $\bm{75.1}$ & $\bm{71.1}$ & $\bm{74.8}$ & \parbox{0pt}{\rule{0pt}{\baselineskip}} \\ [-0.2ex]  		
				dog barking~~~& $74.1$ & $\bm{76.6}$ & $\bm{77.3}$ & $\bm{78.2}$ & $\bm{78.4}$ & \parbox{0pt}{\rule{0pt}{\baselineskip}} \\ [-0.2ex]  		
				footsteps~~~~~ & $45.3$ & $\bm{48.8}$ & $\bm{50.7}$ & $\bm{51.8}$ & $\bm{50.5}$ & \parbox{0pt}{\rule{0pt}{\baselineskip}} \\ [-0.2ex]  	
				glass smash~~~ & $80.1$ & $77.0$ & $\bm{82.0}$ & $\bm{82.6}$ & $\bm{83.9}$ & \parbox{0pt}{\rule{0pt}{\baselineskip}} \\ [-0.2ex]  	
				gun shot~~~~~ & $79.7$ & $67.9$ & $76.3$ & $\bm{82.4}$ & $\bm{83.9}$ & \parbox{0pt}{\rule{0pt}{\baselineskip}} \\ [-0.2ex]  		
				horsewalk~~~~ & $44.7$ & $44.6$ & $\bm{45.6}$ & $44.3$ & $44.2$ & \parbox{0pt}{\rule{0pt}{\baselineskip}} \\ [-0.2ex]  	
				mixer~~~~~~~ & $69.9$ & $\bm{80.9}$ & $\bm{75.9}$ & $\bm{75.5}$ & $\bm{70.9}$ & \parbox{0pt}{\rule{0pt}{\baselineskip}} \\ [-0.2ex]  	
				motorcycle~~~~ & $50.8$ & $43.2$ & $46.1$ & $\bm{55.7}$ & $\bm{55.2}$ & \parbox{0pt}{\rule{0pt}{\baselineskip}} \\ [-0.2ex]   		
				rain~~~~~~~~& $77.9$ & $69.7$ & $77.7$ & $73.9$ & $\bm{82.5}$ & \parbox{0pt}{\rule{0pt}{\baselineskip}} \\ [-0.2ex]  
				thunder~~~~~~ & $60.1$ & $56.9$ & $\bm{62.2}$ & $58.7$ & $\bm{61.3}$ & \parbox{0pt}{\rule{0pt}{\baselineskip}} \\ [-0.2ex]  	
				\cline{1-6}
				{\bf Average~~~} & $60.8$ & $\bm{61.6}$ & $ \bm{63.9}$ & $\bm{64.5}$ & $\bm{64.6}$ & \parbox{0pt}{\rule{0pt}{\baselineskip}} \\ [-0.2ex] 
				{\bf Overall~~~}& $63.1$ & $\bm{64.3}$ & $\bm{66.0}$ & $\bm{66.8}$ & $\bm{65.9}$ & \parbox{0pt}{\rule{0pt}{\baselineskip}} \\ [-0.2ex]  
				
				\cline{1-6}
			\end{tabular}
		\end{center}
		\label{tab:performance}
		\vspace{-0.5cm}
	\end{table}

	\setlength\tabcolsep{2.5pt} 
	\begin{table}[!t]
		\caption{The obtained F1-scores w.r.t. different overlapping degrees. Bold-face indicates the multi-task networks outperform the baseline.}
		\footnotesize
		\vspace{-0.2cm}
		\begin{center}
			\begin{tabular}{|>{\centering\arraybackslash}m{0.75in}|>{\centering\arraybackslash}m{0.4in}|>{\centering\arraybackslash}m{0.4in}|>{\centering\arraybackslash}m{0.4in}|>{\centering\arraybackslash}m{0.4in}|>{\centering\arraybackslash}m{0.45in}|>{\centering\arraybackslash}m{0in} @{}m{0pt}@{}}
				\cline{1-6}
				\multirow{2}{*}{\makecell{Overlapping \\ degree}} & \multirow{2}{*}{\makecell{Multi\\-label}}  &  \multicolumn{4}{c|}{Multi-task} & \parbox{0pt}{\rule{0pt}{0ex+\baselineskip}} \\ [0ex]  	
				\cline{3-6}
				& & 2 tasks & 4 tasks & 8 tasks & 16 tasks & \parbox{0pt}{\rule{0pt}{0ex+\baselineskip}} \\ [-0.2ex] 
				\cline{1-6}
				1 & $70.5$ & $\bm{71.4}$ & $\bm{73.8}$ & $\bm{74.0}$ & $\bm{73.4}$ & \parbox{0pt}{\rule{0pt}{\baselineskip}} \\ [-0.2ex]  	
				2& $56.7$ & $\bm{58.2}$ & $\bm{59.8}$ & $\bm{60.3}$ & $\bm{62.0}$ & \parbox{0pt}{\rule{0pt}{\baselineskip}} \\ [-0.2ex]  	
				3& $48.9$ & $\bm{49.0}$ & $\bm{52.1}$ & $\bm{53.7}$ & $\bm{52.8}$ & \parbox{0pt}{\rule{0pt}{\baselineskip}} \\ [-0.2ex]  
				4& $44.9$ & $42.2$ & $43.7$ & $\bm{48.1}$ & $\bm{47.1}$ &  \parbox{0pt}{\rule{0pt}{\baselineskip}} \\ [-0.2ex]  		
				5& $34.5$ & $\bm{36.1}$ & $\bm{39.9}$ & $\bm{38.0}$ & $\bm{36.0}$ &  \parbox{0pt}{\rule{0pt}{\baselineskip}} \\ [-0.2ex]  	
				6& $26.7$ & $\bm{28.1}$ & $\bm{34.0}$ & $\bm{32.0}$ & $\bm{30.5}$ & \parbox{0pt}{\rule{0pt}{\baselineskip}} \\ [-0.2ex]  	
				\cline{1-6}
			\end{tabular}
		\end{center}
		\label{tab:performance_tut}
		\vspace{-0.5cm}
	\end{table}
	
	\vspace{-0.15cm}
	\subsection{Experimental results}
	\vspace{-0.15cm}
	Given the 16 event categories, there are many possibilities for multi-task decomposition (cf. Section \ref{sec:multitaskdecomposition}). Here, we adopted a simple approach for this purpose. We divided the event categories into $\{2,4,8,16\}$ groups of equal size, which allowed us to study the decompositions of $\{2,4,8,16\}$ tasks. There were $\{8, 4, 2, 1\}$ event categories assigned to each group, resulting in the number of classes $\{256, 16, 4, 2\}$ for each of the tasks, respectively.
	The multi-task decompositions is shown in Table \ref{tab:multitaskdecomposition}. For simplicity, we kept the order of the categories as they appear in the left most column. An optimal task decomposition would be beneficial, however, we leave this open question for future work. Also, it is easy to see that more tasks will result in a larger network and increased computational overhead.
	
	Table \ref{tab:performance} shows the performance obtained by the multi-label baseline and the proposed multi-task networks with respect to different multi-task decompositions. Overall, all the multi-task systems outperform the multi-label baseline, improving the F1-score by $1.2\%$, $2.9\%$, $3.7\%$, and $2.8\%$ absolute with 2, 4, 8, and 16 tasks, respectively. Improvement can also be seen on most of the event categories. The gain is lowest with the 2-task decomposition - that can be explained by the large number of classes (i.e. 256) in the tasks, which is reasonably proportional to the tasks' complexity. Apparently, a decomposition that results in simpler tasks with the small number of classes in each task (for example, 8 tasks with 4 classes each in this case), is more desirable. However, the more tasks resulted from a decomposition, the more task-specific subnets are needed, leading to higher computational cost.  Of note, $N\!=\!Y\!=\!16$, the original task was reduced to 16 one-vs.-rest binary classification problems. However, this was still a multi-task problem, where each of the tasks was handled by a task-specific subnet in a multi-task network. It is different from the multi-label baseline, in which all the inherent one-vs.-rest binary classification problems share the same network.
	
	We further show the performance obtained with respect to different event overlapping degrees in Table \ref{tab:performance_tut}. First, as expected, all the AED systems perform best on monophonic events (i.e., overlapping degree of 1) and the performance decreases monotonically with the increase of overlapping degree, confirming that polyphonic mixtures are the main source of detection errors. Second, the performance gains achieved by the multi-task networks over the multi-label baseline  are distributed quite evenly across all the overlapping degrees, except for the overlapping degree of 4 in cases of 2 and 4 tasks.
	
\begin{figure} [!t]
		\centering
		\includegraphics[width=0.9\linewidth]{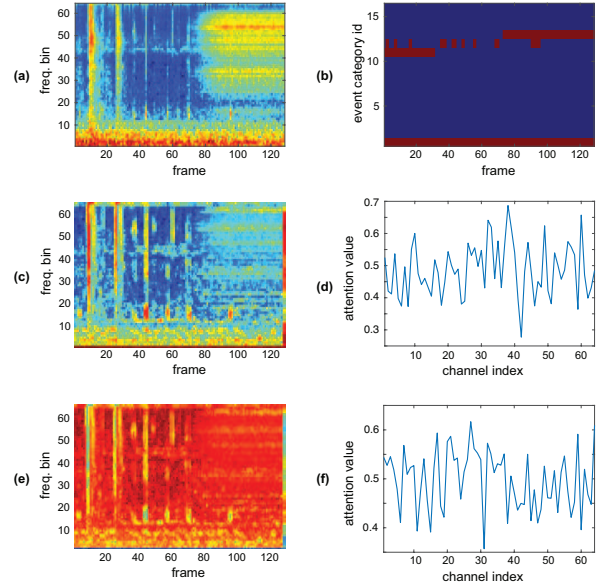}
		\vspace{-0.3cm}
		\caption{\small Task-specific attention masks. (a) log-Mel spectrogram input; (b) event activity; (c) TF attention mask and (d) channel attention mask of \emph{alarm \& siren}; (e) TF attention mask and (f) channel attention mask of \emph{mixer}.}
		\label{fig:attention}
		\vspace{-0.25cm}
	\end{figure}

	
	To gain insight about the task-specific subnets, we visualized in Fig. \ref{fig:attention} their attention masks of the first Att-Conv-Block with respect to an input with the presence of 4 event instances. The multi-task network with 16 tasks was employed for this purpose and the attention masks of the task-specific subnets corresponding to \emph{alarm \& siren} (id=1) and \emph{mixer} (id=13) were selected for visualization. As can be seen from the figure, the attention masks of the two subnets are very distinguishable, suggesting that the subnets learned different TF and channel masks to derive task-specific features from the same backbone feature maps.

	\vspace{-0.15cm}
	\section{Conclusions}
	\label{sec:conclusion}
	\vspace{-0.2cm}
	
	We have framed the AED task as a multi-class classification problem, where all possible event mixtures were considered as the classes. To circumvent the combinatorial explosion, we proposed to divide the set of event categories into multiple non-overlapping groups. In other words, the original problem was reduced to multiple simpler tasks, or a multi-task problem, where a group of event categories was associated with one of the tasks. A multi-task network was then introduced to deal with the multi-task modelling problem. The network was composed of a backbone subnet and multiple task-specific subnets. The backbone was supposed to learn task-shared feature maps while each task-specific subnet was tasked to extract features from the task-shared feature maps and refine them for the task at hand. We demonstrated that the proposed multi-class formulation coupled with the multi-task network was able to detect polyphonic events with varying overlapping degrees more accurate than the commonly used multi-label approach. 
	
	\vspace{-0.15cm}
	\section{Acknowledgement}
	\vspace{-0.15cm}
	H. Phan is supported by a Turing Fellowship under the EPSRC grant EP/N510129/1.
	\vspace{-0.15cm}
	\small
	\bibliographystyle{IEEEbib}
	\bibliography{reference}
\end{document}